\RequirePackage{lineno}
\documentclass[aps,prd, showpacs, twocolumn,superscriptaddress,groupedaddress]{revtex4}  
\usepackage{graphicx}  
\usepackage{dcolumn}   
\usepackage{bm}        
\usepackage{amssymb}   
\usepackage{amsmath}
\usepackage{color} 
\usepackage{array}
\newcommand{\PreserveBackslash}[1]{\let\temp=\\#1\let\\=\temp}
\newcolumntype{C}[1]{>{\PreserveBackslash\centering}p{#1}}
\newcolumntype{R}[1]{>{\PreserveBackslash\raggedleft}p{#1}}
\newcolumntype{L}[1]{>{\PreserveBackslash\raggedright}p{#1}}

\newcommand\ST{\rule[-0.5em]{0pt}{1.5em}} 

\begin{document}
\title{Improved measurements of \bm{$\chi_{cJ} \rightarrow \Sigma^+ \bar{\Sigma}^-$} and \bm{$\Sigma^0 \bar{\Sigma}^0$} decays} 
\author{
  \small
  M.~Ablikim$^{1}$, M.~N.~Achasov$^{9,d}$, S. ~Ahmed$^{14}$, M.~Albrecht$^{4}$, M.~Alekseev$^{53A,53C}$, A.~Amoroso$^{53A,53C}$, F.~F.~An$^{1}$, Q.~An$^{50,40}$, J.~Z.~Bai$^{1}$, Y.~Bai$^{39}$
% ...
\\
\vspace{0.2cm}
(BESIII Collaboration)\\
\vspace{0.2cm} {\it
$^{1}$ Institute of High Energy Physics, Beijing 100049, People's Republic of China\\
% ...
$^{4}$ Bochum Ruhr-University, D-44780 Bochum, Germany\\
% ...
$^{9}$ G.I. Budker Institute of Nuclear Physics SB RAS (BINP), Novosibirsk 630090, Russia\\
% ...
$^{14}$ Helmholtz Institute Mainz, Johann-Joachim-Becher-Weg 45, D-55099 Mainz, Germany\\
% ...
$^{39}$ Southeast University, Nanjing 211100, People's Republic of China\\
$^{40}$ State Key Laboratory of Particle Detection and Electronics, Beijing 100049, Hefei 230026, People's Republic of China\\
% ...
$^{50}$ University of Science and Technology of China, Hefei 230026, People's Republic of China\\
% ...
$^{53}$ (A)University of Turin, I-10125, Turin, Italy; (B)University of Eastern Piedmont, I-15121, Alessandria, Italy; (C)INFN, I-10125, Turin, Italy\\
% ...
$^{d}$ Also at the Novosibirsk State University, Novosibirsk, 630090, Russia\\
% ...
}
}

\date{\today}

\begin{abstract}
Using a data sample of $(448.1\pm2.9)\times 10^6~\psi(3686)$ events collected with the BESIII detector at the BEPCII collider, we present measurements of branching fractions for the decays $\chi_{cJ} \rightarrow \Sigma^+ \bar{\Sigma}^-$ and $\Sigma^0 \bar{\Sigma}^0$. The decays $\chi_{c1,2}\rightarrow \Sigma^+ \bar{\Sigma}^-$ and $\Sigma^0 \bar{\Sigma}^0$ are observed for the first time, and the branching fractions for $\chi_{c0}\rightarrow \Sigma^+ \bar{\Sigma}^-$ and $\Sigma^0 \bar{\Sigma}^0$ decays are measured with improved precision. The branching fraction ratios between the charged and neutral modes are consistent with the prediction of isospin symmetry. 
\end{abstract}
\maketitle

\section{Introduction}
Experimental studies of charmonium decays can test calculations in quantum chromodynamics (QCD) and QCD based effective field theories.
Contributions of the color octet mechanism (COM)~\cite{Bodwin95} to decays of $P$-wave heavy quarkonia have been proposed for more than two decades, and many theoretical predictions for exclusive $\chi_{cJ}$ decays to baryon anti-baryon pairs~\cite{Ping2005, Liu2011, Wong00} have been made. However, there are large differences between predictions and the experimental measurements, \emph{e.g.}, the branching fractions (BF) of $\chi_{c0} \rightarrow \Sigma^+ \bar{\Sigma}^-$ and $\Sigma^0 \bar{\Sigma}^0$ decays as measured by CLEO-c  \cite{CLEOcc} and BESIII \cite{proton} are observed to violate the helicity selection rule from perturbative QCD (pQCD)~\cite{Brodsky1981, Chernyak82, Chernyak84} and also do not agree with models based on the charm meson loop mechanism~\cite{Liu2011, YJzhang, XHLiu}. Further tests of the COM using more decay channels are thus an important input for the development of the theoretical models. 
  
The $\chi_{cJ}$ ($J=0, 1, 2$) states are identified as the charmonium $P$-wave spin triplet. Although they cannot be produced directly in the annihilation of electrons with positrons, the radiative decays of the $\psi(3686)$ meson can generate large numbers of these particles. In this article, measurements of the BF of $\chi_{cJ} \rightarrow \Sigma^+ \bar{\Sigma}^-$ and $\Sigma^0 \bar{\Sigma}^0$ decays are presented using the world's largest statistics of $N_{\psi(3686)}=(448.1 \pm 2.9) \times 10^6 ~\psi(3686)$ events~\cite{Zhiyong17} at on-threshold production collected with the BESIII detector. In addition, the isospin symmetry is tested using the BF ratios between the charged and neutral modes. 

\section{BESIII DETECTOR AND Monte Carlo SIMULATION}
\label{sec:BES}
The BESIII experiment is operated at the Beijing electron positron collider II (BEPCII), which reaches a peak luminosity of $1.0\times10^{33}~\text{cm}^{-2}\text{s}^{-1}$ at a center-of-mass energy of 3773~MeV. The detector has a geometrical acceptance of 93\% of the solid angle and is comprised of four main components. A helium-gas based main drift chamber (MDC) is used to track charged particles. The single wire resolution is better than 130~$\mu$m, which, together with a magnetic field of 1~T leads to a momentum resolution of 0.5\% at 1~GeV/$c$. The energy loss per path length $dE/dx$ is measured with a resolution of 6\%. The MDC is surrounded by a time-of-flight (TOF) system built from plastic scintillators. It provides a $2\sigma ~K/\pi$ separation up to 1~GeV/$c$ momentum with a time resolution of 80 (110) ps for the barrel (end-caps). Particle energies are measured in the CsI(Tl) electro-magnetic calorimeter (EMC), which achieves an energy resolution for electrons of 2.5\% (5\%) at 1~GeV/$c$ momentum and a position resolution of 6~mm (9~mm) for the the barrel (end-caps). Outside of the magnet coil, a muon detector based on resistive plate chambers (RPC) provides a spatial resolution of better than 2~cm. A more detailed description of the detector can be found in Ref.~\cite{NIM10}. 

A \textsc{Geant4}~\cite{Geant4} based Monte Carlo (MC) simulation package is used to optimize the event selection, estimate the signal efficiency and the background level. The event generator KKMC~\cite{Jadach01} simulates the electron-positron annihilation and the production of the $\psi$ resonances. Particle decays are generated by EVTGEN~\cite{Lange01} for the known decay modes with BFs taken from the Particle Data Group (PDG)~\cite{PDG16} and LUNDCHARM~\cite{Lundcharm00} for the unknown ones. A generic MC sample containing all possible decay channels is used to study backgrounds, while  signal MC samples containing only the exclusive decay channels are used to determine efficiencies. 
In the signal MC simulation, the decay $\psi(3686) \rightarrow \gamma \chi_{cJ}$ is generated according to the angular distributions from Ref.~\cite{Tanenbaum78}, where the photon polar angle $\theta$ is distributed according to $(1+\text{cos}^2\theta)$, $(1-\frac{1}{3}\text{cos}^2\theta)$, $(1+\frac{1}{13}\text{cos}^2\theta)$ for $\psi(3686) \rightarrow \gamma \chi_{c0, 1, 2}$ decays, respectively. The decays $\chi_{cJ}$ to baryon anti-baryon pairs are generated with the phase space model, and the weak decays of baryons are generated with a model taking into account parity violation. 

\begin{figure*}[!t]
  \includegraphics[width=0.85\columnwidth]{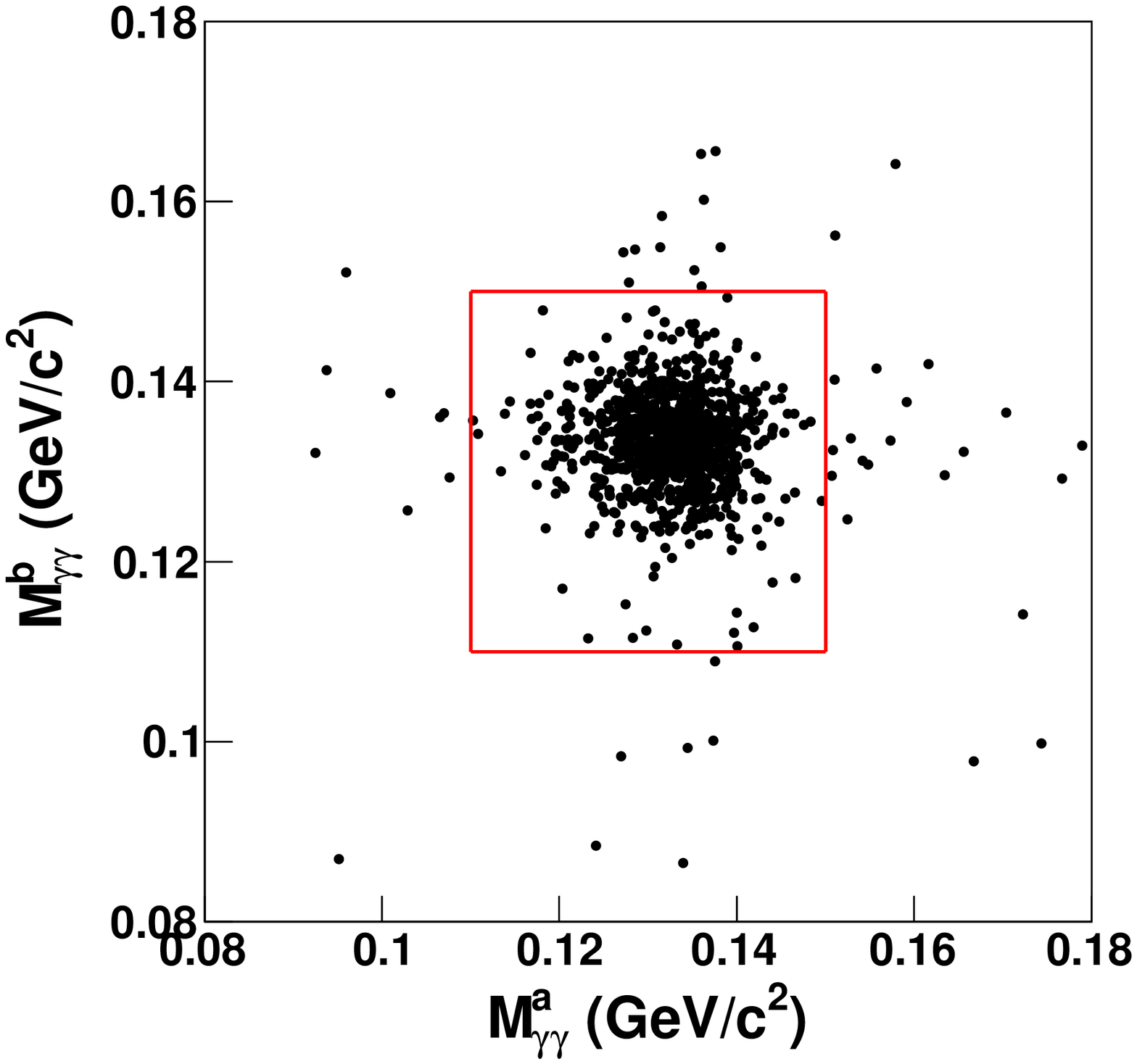}
   \includegraphics[width=0.85\columnwidth]{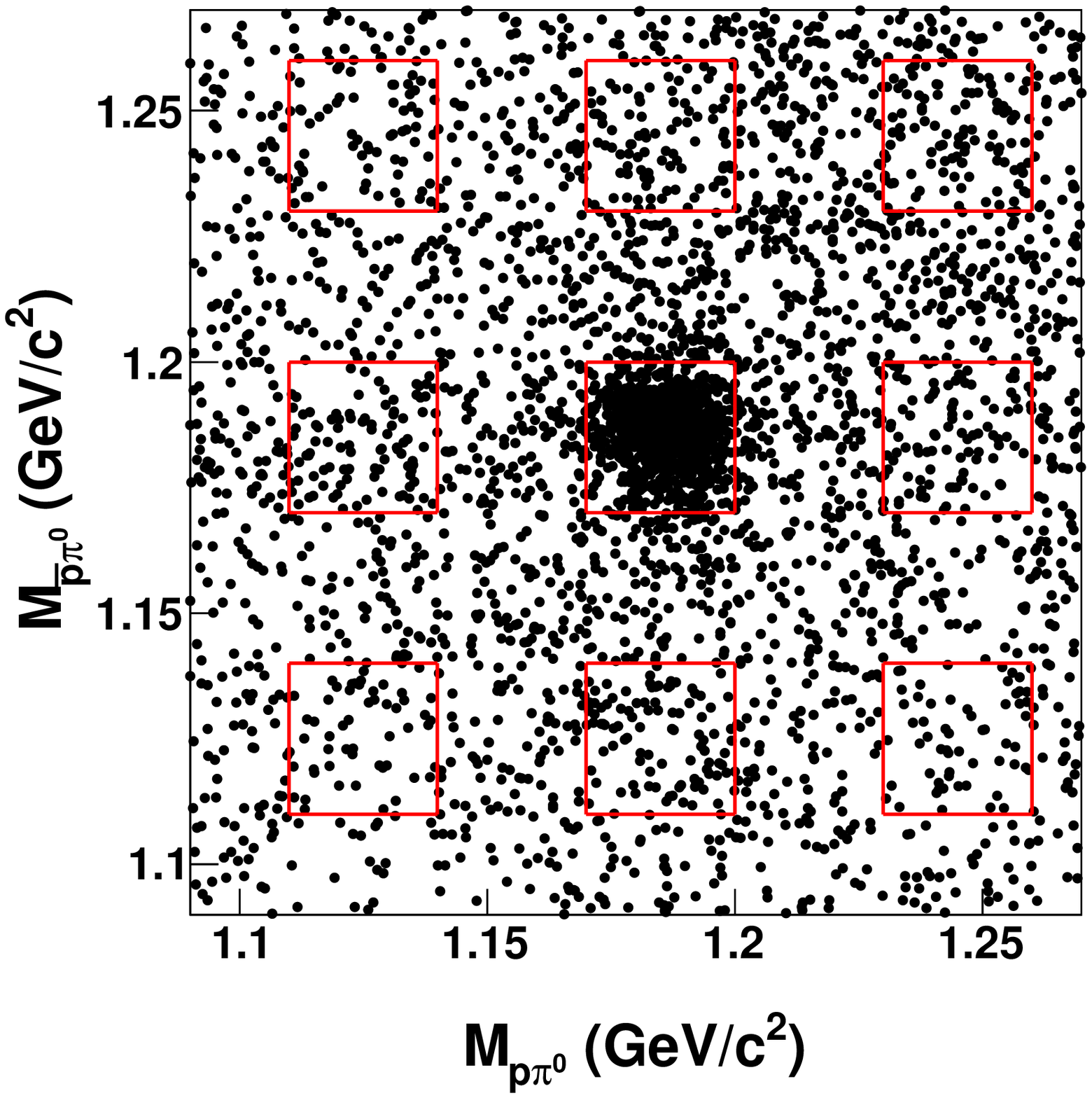}
   \renewcommand{\figurename}{Fig.}
   \caption{Distribution of $M_{\gamma\gamma}^{a}$ versus $M_{\gamma\gamma}^{b}$ (left)  and distribution of $M_{p\pi^0}$ versus $M_{\bar{p}\pi^0}$ (right) for $\chi_{cJ} \rightarrow \Sigma^+ \bar{\Sigma}^-$. The central (surrounding) boxes indicate the signal (sideband) regions. }
   \label{fig:scatt}
 \end{figure*}

\section{Event selection}
\label{sec:measurement}
 \subsection{\boldmath $\chi_{cJ} \rightarrow \Sigma^+ \bar{\Sigma}^-$}

In the decay chain $\psi(3686) \rightarrow \gamma \chi_{cJ},\quad \chi_{cJ} \rightarrow \Sigma^+ \bar{\Sigma}^-$, the $\Sigma^+$ $(\bar{\Sigma}^-)$ particle is
reconstructed in the decay channel $p \pi^0$  $(\bar{p} \pi^0), \, \pi^0 \rightarrow \gamma\gamma$. Thus at least five photons and two charged tracks with zero net charge are required in the final state. 
Charged tracks are selected by requiring a value of the polar angle $|\cos\theta|$ of less than 0.93 and a point of closest approach to the nominal interaction point within 15~cm in beam direction ($V_z$) and within 2~cm in the plane transverse to the beams ($V_r$). Larger requirements on $V_z$ and $V_r$ are used compared to the nominal cuts ($V_z \le 10~\text{cm}, V_r \le 1~\text{cm}$) due to the decay length of  $\Sigma^+ ~(\bar{\Sigma}^-)$ particle. The $dE/dx$ information obtained from the MDC and time information from the TOF system is combined in a global likelihood to identify protons and anti-protons. The (anti-)proton likelihood is required to be larger than the one obtained with a pion and kaon hypothesis.
Photon candidates are reconstructed from EMC showers and are required to have an energy of greater than 25~MeV for the barrel ($|\cos\theta|<0.8$) or greater than 50~MeV in the end-cap regions ($0.86<|\cos\theta|<0.92$). In addition, the timing of good photon candidates is required to be within 700~ns of the collision event, in order to reduce contributions from electronics noise and beam-related background. 
A four-constraint (4C) kinematic fit is applied using the $\psi(3686) \rightarrow 5 \gamma p \bar{p}$ hypothesis. In events with more than five photon candidates, the combination with the least $\chi^2_{4C}$ is chosen for further analysis.  The $\chi^2_{4C}$ is required to be less than 50.  The $\pi^0$ candidates are reconstructed by minimizing $\sqrt{(M_{\gamma\gamma}^{a}-M_{\pi^0})^2+(M_{\gamma\gamma}^{b}-M_{\pi^0})^2}$, where $M_{\gamma\gamma}^{a, b}$ and $M_{\pi^0}$ represent the invariant mass of $\gamma \gamma$ pairs and the nominal $\pi^0$ mass, respectively.  The reconstructed $\pi^0$ mass is required to be in the range from 0.11 to 0.15~GeV/$c^2$. The left panel of Fig.~\ref{fig:scatt} shows the distribution of $M_{\gamma\gamma}^{a}$ versus $M_{\gamma\gamma}^{b}$ in data.  The $\Sigma^+$ and $\bar{\Sigma}^-$ baryons are reconstructed by minimizing $\sqrt{(M_{p\pi^0}-M_{\Sigma^+})^2+(M_{\bar{p}\pi^0}-M_{\bar{\Sigma}^-})^2}$, where $M_{p\pi^0}$ ($M_{\bar{p}\pi^0}$) and  $M_{\Sigma^+}$($M_{\bar{\Sigma}^-}$) represent the invariant mass of $p\pi^0$ ($\bar{p}\pi^0$) and nominal $\Sigma^+$ ($\bar{\Sigma}^-$) mass, respectively. The reconstructed masses of the $\Sigma^+$ and the $\bar{\Sigma}^-$ particles are required to fall into the interval 1.17-1.20~GeV/$c^2$.
The probability of assigning photons to the wrong $\pi^0$ and the wrong $\pi^0$ to a $\Sigma^+/\bar{\Sigma}^-$ particle is studied using the signal MC sample and found to be lower than 0.5\% and 0.1\%, respectively. The right panel of Fig.~\ref{fig:scatt} shows the distribution of  $M_{p\pi^0}$ versus $M_{\bar{p}\pi^0}$ in data. To remove the $\psi(3686) \rightarrow \Sigma^+ \bar{\Sigma}^-$ background, the invariant mass of $\Sigma^+ \bar{\Sigma}^-$ is required to be below 3.6~GeV$/c^2$. 

\begin{figure}[!t]
  \includegraphics[width=0.9\columnwidth]{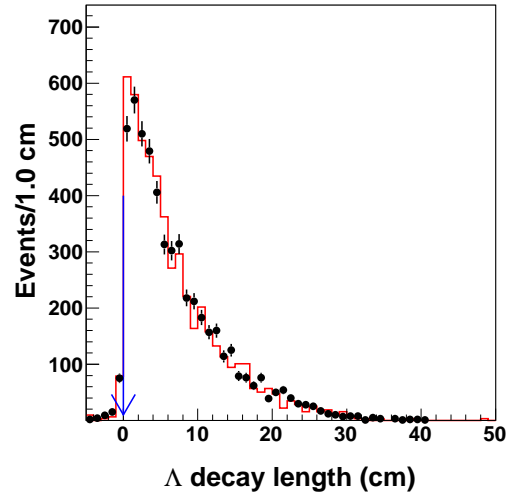}
   \renewcommand{\figurename}{Fig.}
   \caption{The distribution of the $\Lambda$ decay length for data (points) and MC (histogram). }
   \label{fig:lambdel}
 \end{figure}

\begin{figure*}[!t]
  \includegraphics[width=0.85\columnwidth]{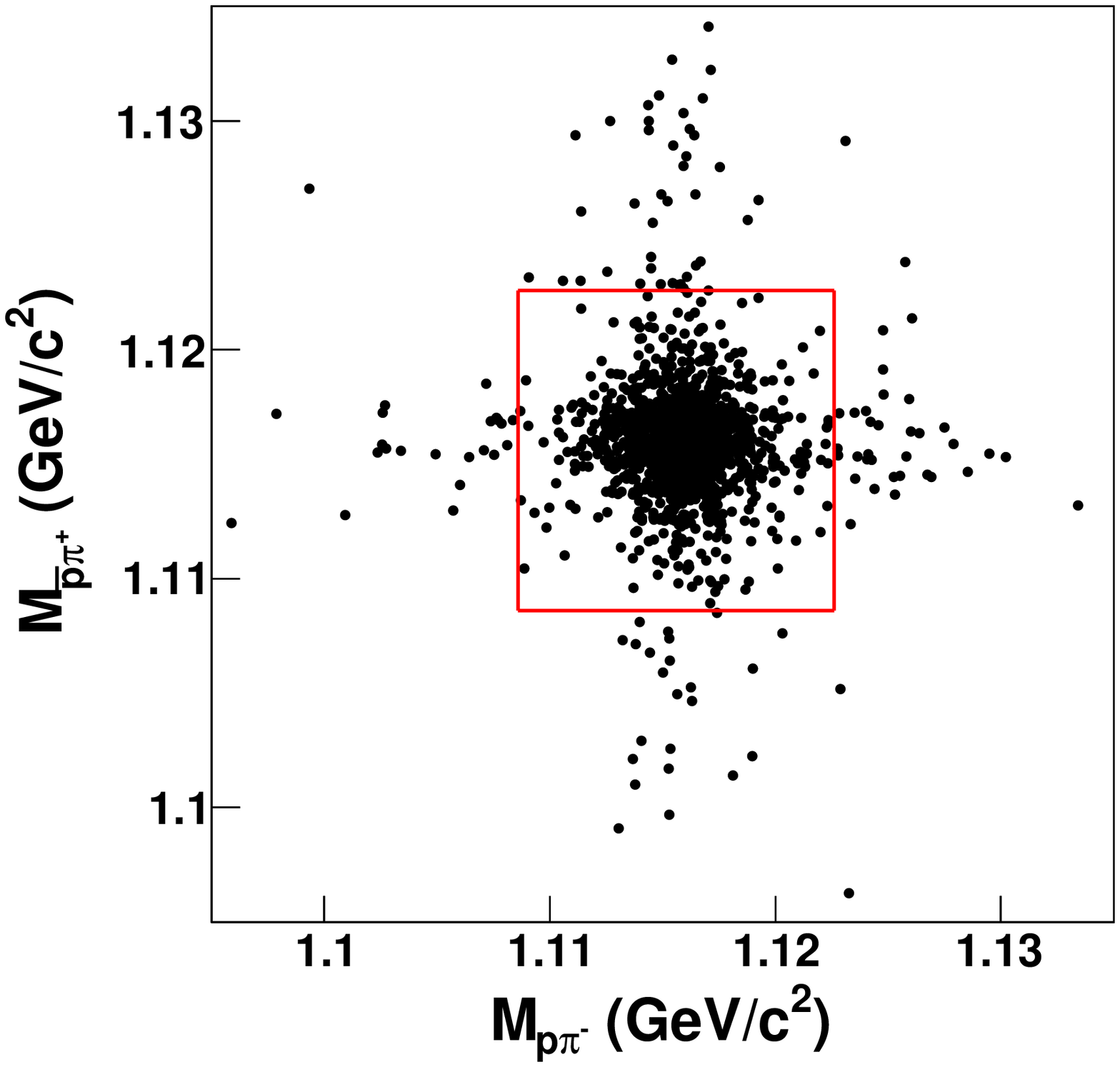}
   \includegraphics[width=0.85\columnwidth]{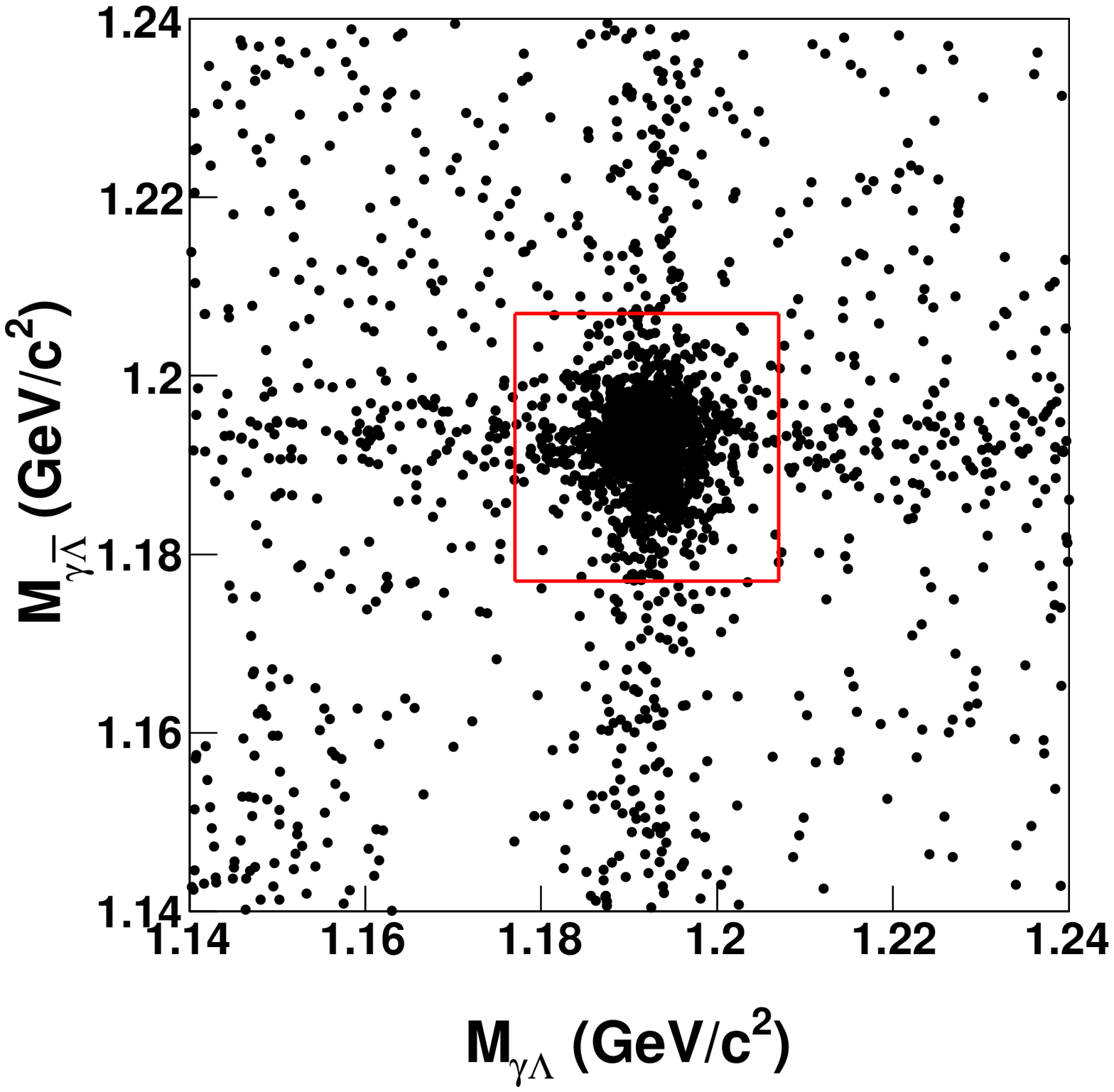}
   \renewcommand{\figurename}{Fig.}
   \caption{Distribution of $M_{p \pi^-}$ versus $M_{\bar{p} \pi^+}$ (left)  and distribution of $M_{\gamma\Lambda}$ versus $M_{\gamma\bar{\Lambda}}$ (right) for $\chi_{cJ} \rightarrow \Sigma^0 \bar{\Sigma}^0$. The solid boxes indicate the signal regions.} 
   \label{fig:scatt0}
 \end{figure*}
   
\subsection{\boldmath $\chi_{cJ} \rightarrow \Sigma^0 \bar{\Sigma}^0$}
In the decay chain $\psi(3686) \rightarrow \gamma \chi_{cJ},\quad \chi_{cJ} \rightarrow \Sigma^0 \bar{\Sigma}^0$, the $\Sigma^0$ $(\bar{\Sigma}^0)$ particle is reconstructed in the decay channel $\gamma \Lambda$ $(\gamma \bar{\Lambda}),\, \Lambda \rightarrow p \pi^- ~(\bar{\Lambda} \rightarrow \bar{p} \pi^+)$. At least three photons and four charged tracks with zero net charge are required in the event. The selection of charged tracks and good photons are the same as for the $\chi_{cJ} \rightarrow \Sigma^+ \bar{\Sigma}^-$ channel, except that no requirements are placed on the point of closest approach for the tracks since the $\Lambda$ baryon has a large decay length of $c\tau=7.9$~cm. A vertex fit is performed to pairs of charged tracks and a second vertex fit is then performed to the reconstructed $\Lambda$ and $\bar{\Lambda}$ candidates with the requirement of a common point of origin. The signed decay lengths of the $\Lambda$ and the $\bar{\Lambda}$ particle are required to be greater than 0. Figure~\ref{fig:lambdel} shows the distribution of the $\Lambda$ decay length for data and simulation.  The reconstructed invariant masses of the $\Lambda$ and $\bar{\Lambda}$ candidates are required to be within $\pm7$~MeV/$c^2$ of the nominal mass. The left panel of Fig.~\ref{fig:scatt0} shows the distribution of $M_{p \pi^-}$ versus $M_{\bar{p} \pi^+}$ in data, where  $M_{p \pi^-}$ and $M_{\bar{p} \pi^+}$ represent the invariant mass of $p \pi^-$ and $\bar{p} \pi^+$, respectively. A 4C kinematic fit under the hypothesis of the $\psi(3686) \rightarrow 3 \gamma \Lambda \bar{\Lambda}$ decay is applied, imposing energy and momentum conservation. For events with more than three photon candidates, the combination with the least $\chi^2_{4C}$ is kept for further analysis. The $\chi^2_{4C}$ is required to be less than 30. The $\Sigma^0$ and $\bar{\Sigma}^0$ particles are selected by minimizing $\sqrt{(M_{\gamma \Lambda}-M_{\Sigma^0})^2 + (M_{\gamma \bar{\Lambda}}-M_{\bar{\Sigma}^0})^2}$, where $M_{\gamma \Lambda}$ ($M_{\gamma \bar{\Lambda}}$) and $M_{\Sigma^0}$ ($M_{\bar{\Sigma}^0}$) represent the invariant mass of $\gamma \Lambda$ ($\gamma \bar{\Lambda}$) and the nominal $\Sigma^0$ ($\bar{\Sigma}^0$) mass, respectively. The reconstructed $\Sigma^0$ and $\bar{\Sigma}^0$ mass is required to lie in a window of $\pm15$~MeV/$c^2$ around the nominal mass.  The probability of assigning wrong photons in the reconstruction of the $\Sigma^0$ and $\bar{\Sigma}^0$ particle is studied using signal MC and found to be lower than 0.2\%. The right panel of Fig.~\ref{fig:scatt0} shows the distribution of $M_{\gamma\Lambda}$ versus $M_{\gamma\bar{\Lambda}}$ in data. To remove the $\psi(3686) \rightarrow \Sigma^0 \bar{\Sigma}^0$ background, the invariant mass of $\Sigma^0 \bar{\Sigma}^0$ is required to be below 3.6~GeV$/c^2$. 
 
\section{Background study}
\label{sec:bkgstd} 
 \begin{figure*}[!t]
    \includegraphics[width=0.95\columnwidth]{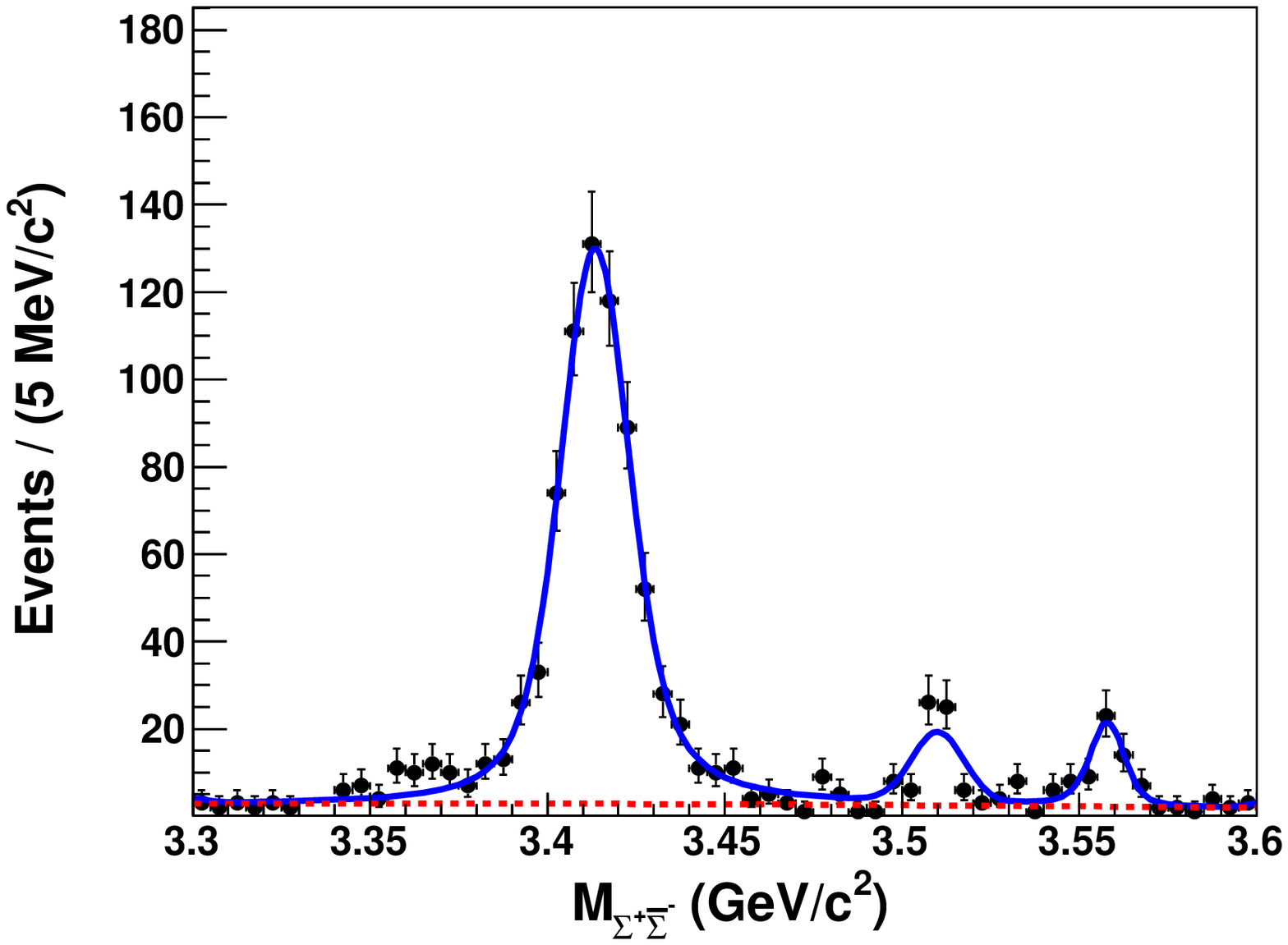}
    \includegraphics[width=0.95\columnwidth]{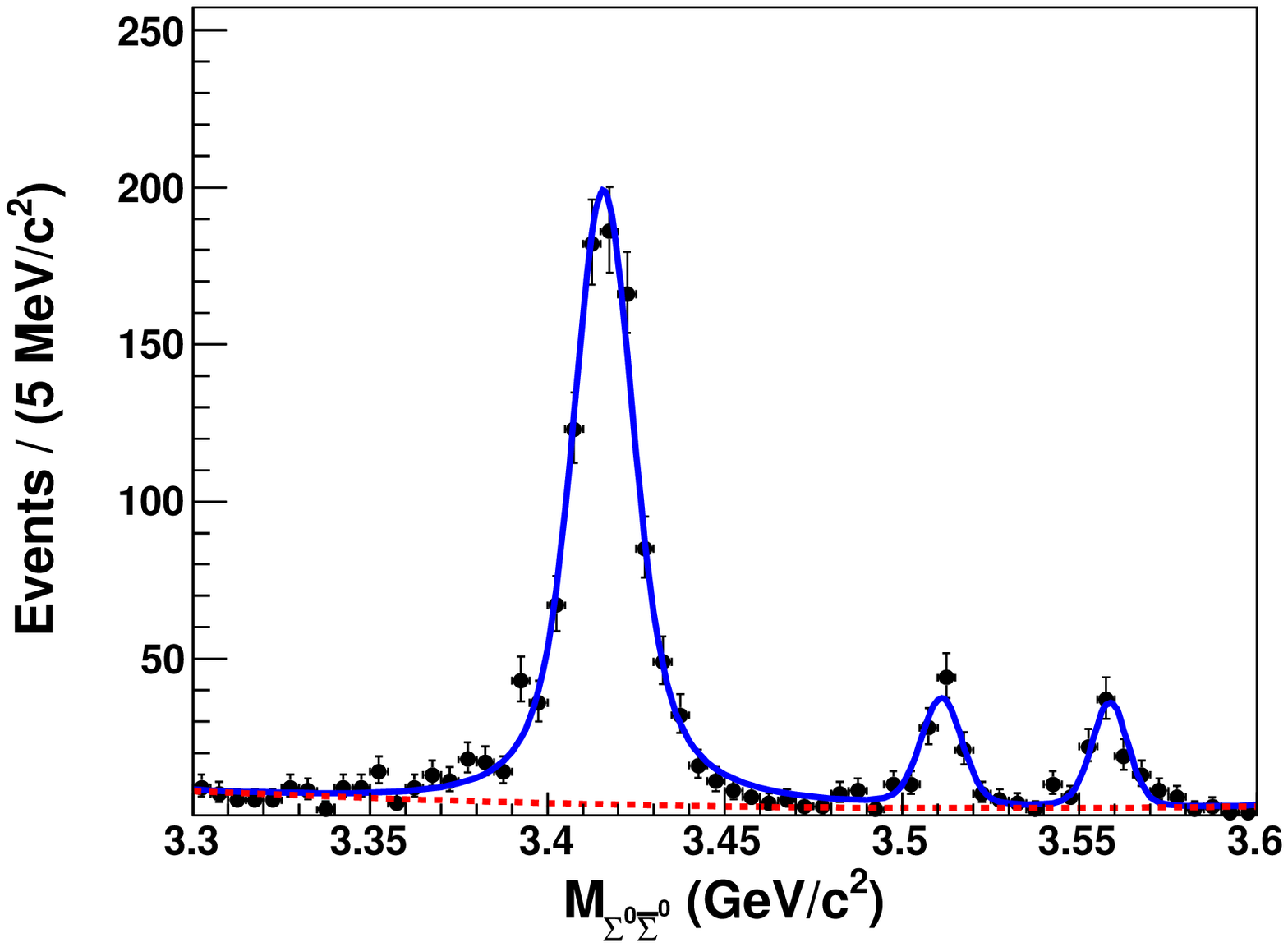}
   \renewcommand{\figurename}{Fig.}
   \caption{Fit results to the invariant mass spectra of $\Sigma^+ \bar{\Sigma}^-$ (left) and $\Sigma^0 \bar{\Sigma}^0$ (right). The dots with error bars represent the data, the solid line represents the fit results and the dashed line represents the smooth background.}
   \label{fig:fitresults}
 \end{figure*}
 
Background from continuum quantum electrodynamics (QED) processes, cosmic rays, beam-gas and beam-wall interactions is estimated using the data collected outside of the $\psi(3686)$ peak. The estimated background is less than 4.4 and 6.3 events  for the $\chi_{cJ} \rightarrow \Sigma^+ \bar{\Sigma}^-$ and $\Sigma^0 \bar{\Sigma}^0$ decay, respectively. 
 
A potential peaking background to the decay $\chi_{cJ} \rightarrow \Sigma^+ \bar{\Sigma}^-$ is the decay $\chi_{cJ} \rightarrow p \bar{p} \pi^0 \pi^0$ without intermediate resonances. We study this peaking background using the two-dimensional sidebands in $M_{p\pi^0}$ versus $M_{\bar{p}\pi^0}$ as shown by the eight surrounding boxes in right panels of Fig.~\ref{fig:scatt}.  The scaling of the sidebands to the signal region is estimated using a phase space distributed MC sample of the process  $\chi_{c0} \rightarrow p \bar{p} \pi^0 \pi^0$, where the scale factor $s$ is obtained by the number of events in the signal region divided by that in each sideband region. After obtaining the invariant mass distribution of $p \bar{p} \pi^0 \pi^0$ from the sidebands, the $\chi_{cJ}$ shape is parametrized with a Breit-Wigner function (BW) convoluted with a Gaussian function, and the background is parametrized with a second-order Chebyshev polynomial. The number of peaking background events $N_{\text{peaking}}$ for the $\chi_{c0}, \chi_{c1}$ and $\chi_{c2}$ signals is estimated to be $20.2 \pm 2.4$, $3.8 \pm 1.4$, and  $7.4 \pm 1.8$, respectively, where the uncertainties are statistical only. A similar study is performed for the $\chi_{cJ} \rightarrow \Sigma^0 \bar{\Sigma}^0$ decays, and no significant peaking background is found. 

The main contributions to non-peaking background for $\chi_{cJ} \rightarrow \Sigma^+ \bar{\Sigma}^-$ are the decays $\psi(3686) \rightarrow \gamma \Sigma^+ \bar{\Sigma}^-$ without the intermediate $\chi_{cJ}$ state, $\psi(3686) \rightarrow \pi^0 \Sigma^+ \bar{\Sigma}^-$,  $\psi(3686) \rightarrow \Sigma^+ \bar{\Sigma}^-$ and non-$\Sigma^+ \bar{\Sigma}^-$ background (mainly $\psi(3686) \rightarrow \pi^0 \pi^0 J/\psi, \quad J/\psi \rightarrow \gamma p \bar{p}$, $J/\psi \rightarrow \pi^0 p \bar{p}$ or $J/\psi \rightarrow p \bar{p}$) from $\psi(3686)$ decays. The background from  $\psi(3686) \rightarrow \Sigma^+ \bar{\Sigma}^-$ decay lies in the $\psi(3686)$ mass region, and can easily be removed by requiring the invariant mass of the $\Sigma^+ \bar{\Sigma}^-$ pair to be below 3.6~GeV/$c^2$. The backgrounds for the decay of $\chi_{cJ} \rightarrow \Sigma^0 \bar{\Sigma}^0$ are similar, replacing the charged with the neutral decay modes.  In addition, there is background from $\psi(3686)\rightarrow \Sigma^0 \pi^0 \bar{\Lambda} + c.c.$ and $\bar{\Sigma}^0 \gamma \Lambda + c.c.$, which contributes to the horizontal and vertical bands around the $\Sigma^0/\bar{\Sigma}^0$ mass region. All non-peaking backgrounds including the QED contribution are found to be smoothly distributed under the $\chi_{cJ}$ peaks and can be modeled by a polynomial function. 

\section{Determination of the $\chi_{cJ}$ signals}
\label{sec:measurementa}
To determine the number of $\chi_{cJ} \rightarrow \Sigma^+ \bar{\Sigma}^-$ events, an extended
unbinned maximum-likelihood fit is performed to the $\Sigma^+ \bar{\Sigma}^-$ invariant mass
distribution between 3.3 and 3.6~GeV/$c^2$.  The $\chi_{cJ}$ signal peaks are described by
probability density functions
\begin{equation}
\label{eq:pdf}
F_J(m)=(BW_J(m) \times E_{\gamma}^{3} \times D(E_{\gamma}))\otimes G(0; \sigma_{\text{res},J}),
\end{equation}
where $BW_J(m)$ is a Breit-Wigner function; $G(0;
\sigma_{\text{res},J})$ is a Gaussian function with the mean value of
zero and a standard deviation of the detection resolution $\sigma_{\text{res},J}$; $E_{\gamma}^3$ is
the cube of radiative photon energy reflecting the energy dependence of the electric dipole (E1)
matrix element; $D(E_{\gamma})$ is a damping factor needed to suppress the diverging tail caused by
the $E_{\gamma}^3$ dependence and is given by $e^{-\frac{E_{\gamma}^2}{8\beta^2}}$, with
$\beta=65$~MeV as determined by the CLEO collaboration~\cite{CLEO}.  The background is described by
a second-order Chebychev polynomial function.  In the fit, the signal yields and the masses of all
three $\chi_{cJ}$ signals as well as the width of the $\chi_{c0}$ signal are left free, while the
detection resolution and the width of the $\chi_{c1}$ and $\chi_{c2}$ resonances are fixed.

The left panel of Fig.~\ref{fig:fitresults} shows the fit result; to estimate the goodness-of-fit, the reduced $\chi^2$ value is determined to be $\chi^2/ndf=73.2/52$.  The statistical significances of the $\chi_{c1, 2} \rightarrow \Sigma^+ \bar{\Sigma}^-$ signal are 8.7~$\sigma$ and 7.1~$\sigma$, respectively.  The statistical significance of the signal is calculated using the changes in the log-likelihood values and the corresponding change in the number of degrees of freedom with and without the signal channel in the fit. A similar fit is performed to the $\Sigma^0 \bar{\Sigma}^0$ invariant mass distribution as shown in the right panel of Fig.~\ref{fig:fitresults}, for which the goodness-of-fit is estimated as $\chi^2/ndf=76.6/52$. The statistical significances for the $\chi_{c1, 2} \rightarrow \Sigma^0 \bar{\Sigma}^0$ decay are 11.8~$\sigma$ and 10.9~$\sigma$, respectively.
 Table~\ref{tab:fitres} lists the detection efficiencies obtained from MC simulation and the numbers of observed events for the $\chi_{cJ}$ signals.  To calculate the efficiency for the decay $\chi_{cJ} \rightarrow \Sigma^+ \bar{\Sigma}^- $, the track helix parameters for the proton and anti-proton are corrected in simulation (as described in \cite{helix} in detail) to improve the consistency of the 4C kinematic fit between data and MC simulation, where the correction factors are obtained using a
control sample of $\psi(3686) \rightarrow \pi^0 \pi^0 J/\psi; J/\psi \rightarrow p \bar{p} \gamma$ decay. 

 \begin{table}[thbp]
  \caption[fitted results]{The detection efficiency ($\epsilon$) obtained from MC simulation and the number of observed events for $\chi_{cJ}$ signal ($N_{\text{obs}}$). The uncertainty is statistical only.} 
  \label{tab:fitres}
  \begin{tabular}{lccccc} \hline
    \hline       Decay channel~~                                &  ~~~$\epsilon$ (\%)~~~~~     & ~~$N_{\text{obs}}$  \\ 
    \hline       $\chi_{c0} \rightarrow \Sigma^+ \bar{\Sigma}^- $  & $12.95\pm0.05$ & $747.4\pm35.4$  \ST \\
                    $\chi_{c1} \rightarrow \Sigma^+ \bar{\Sigma}^- $  & $14.03\pm0.05$ & $58.9\pm9.4$ \\
                    $\chi_{c2} \rightarrow \Sigma^+ \bar{\Sigma}^- $  & $13.18\pm0.05$ & $54.7\pm9.3$ \\
                    $\chi_{c0} \rightarrow \Sigma^0 \bar{\Sigma}^0 $  & $12.19\pm0.05$ & $1045.8\pm40.1$ \\
                    $\chi_{c1} \rightarrow \Sigma^0 \bar{\Sigma}^0 $  & $13.46\pm0.05$ & $103.2\pm11.9$ \\
                    $\chi_{c2} \rightarrow \Sigma^0 \bar{\Sigma}^0$  & $13.07\pm0.05$ & $90.8\pm11.7$ \\
                    \hline
    \hline
  \end{tabular}
\end{table}

\begin{table*}[thbp]
  \caption[BF results]{The BF results for the measurement of $\chi_{cJ} \rightarrow \Sigma^+ \bar{\Sigma}^-$ and $\Sigma^0 \bar{\Sigma}^0$ (second column), together with values from PDG world average~\cite{PDG16}, previous measurement from BESIII publications~\cite{proton}, CLEO~\cite{CLEOcc} and theoretical predictions~\cite{Ping2005, Liu2011, Wong00} for comparison. To make an objective comparison, the BF of $\chi_{cJ} \rightarrow \Sigma \bar{\Sigma}$ decays from previous BESIII are corrected with the newest BF of $\psi(3686) \rightarrow \gamma \chi_{cJ}$ from Ref.~\cite{PDG16}. To be independent of the BF of $\psi(3686) \rightarrow \gamma \chi_{cJ}$, the product BF ($\mathcal{B}_{\rm prod}$) of  $\psi(3686) \rightarrow \gamma \chi_{cJ}$ and $\chi_{cJ} \rightarrow \Sigma \bar{\Sigma}$ are also listed (last column). The first uncertainty is statistical and the second systematic. Throughout the table, the BFs are given in units of $10^{-5}$.}
  \label{tab:finalres}
  \begin{tabular}{lcccccc} \hline
   \hline    Channel   &  This work  & PDG &  Previous BESIII~\cite{proton} & CLEO~\cite{CLEOcc} & Theory  & $\mathcal{B}_{\rm prod}$\\ 
    \hline   
                $\chi_{c0} \rightarrow \Sigma^+ \bar{\Sigma}^-$  & $50.4\pm2.5\pm2.7$ &  $39\pm7$ & $43.7\pm4.0\pm2.8$ & $32.5\pm5.7\pm4.3$ & 5.5-6.9~\cite{Liu2011}  & $4.99\pm0.24\pm0.24$ \ST \\
                 $\chi_{c1} \rightarrow \Sigma^+ \bar{\Sigma}^-$  & $3.7\pm0.6\pm0.2$ &  $<6$  & $5.2\pm1.3\pm0.5 (<8.3)$ & $<6.5$ & 3.3~\cite{Wong00} & $0.35\pm0.06\pm0.02$ \\
                 $\chi_{c2} \rightarrow \Sigma^+ \bar{\Sigma}^-$  & $3.5\pm0.7\pm0.3$ &  $<7$  & $4.7\pm1.8\pm0.7 (<8.4)$ & $<6.7$ &5.0~\cite{Wong00} & $0.32\pm0.06\pm0.03$\\
                 $\chi_{c0} \rightarrow \Sigma^0 \bar{\Sigma}^0$  & $47.7\pm1.8\pm3.5$ &  $44\pm4$ & $46.0\pm3.3\pm3.7$ & $44.1\pm5.6\pm4.7$ & ($25.1\pm3.4$, $18.7\pm4.5$)~\cite{Ping2005} & $4.72\pm0.18\pm0.28$\\
                 $\chi_{c1} \rightarrow \Sigma^0 \bar{\Sigma}^0$  & $4.3\pm0.5\pm0.3$ &  $<4$  & $3.7\pm1.0\pm0.5 (<6.0)$ & $<4.4$ & 3.3~\cite{Wong00} & $0.41\pm0.05\pm0.03$\\
                 $\chi_{c2} \rightarrow \Sigma^0 \bar{\Sigma}^0$  & $3.9\pm0.5\pm0.3$ &  $<6$  & $3.8\pm1.0\pm0.5 (<6.2)$ & $<7.5$ & ($38.9\pm8.8$, $4.2\pm0.5$)~\cite{Ping2005}  & $0.35\pm0.05\pm0.02$ \\
                ~ & ~ &  ~  & ~ & ~ &  5.0~\cite{Wong00} & ~  \\ \hline
    \hline
  \end{tabular}
\end{table*}

Using the quantities listed in Table~\ref{tab:fitres} and the BF ($\mathcal{B}_j$) of the intermediate states obtained from the PDG~\cite{PDG16}, the BF ($\cal{B}$) of $\chi_{cJ} \rightarrow \Sigma^+ \bar{\Sigma}^-$ and $\Sigma^0 \bar{\Sigma}^0$ decays are calculated by
\begin{equation}
\label{eq:calfunction}
\mathcal{B}=\frac{N_{\text{obs}}-N_{\text{peaking}}} {N_{\psi(3686)}\times \epsilon \times \prod_j \mathcal{B}_j }.
\end{equation}
The results are listed in Table~\ref{tab:finalres}, together with the values from theoretical predictions~\cite{Ping2005, Liu2011, Wong00}, previous measurement from BESIII~\cite{proton}, CLEO~\cite{CLEOcc} and the PDG world averages~\cite{PDG16} for comparison.  Note that we use the prediction of the decay $\chi_{c0} \rightarrow \Sigma^- \bar{\Sigma}^+$ from Ref.~\cite{Liu2011}  for $\chi_{c0} \rightarrow \Sigma^+ \bar{\Sigma}^-$ due to isospin symmetry.   The previous results on $\chi_{c1, 2} \rightarrow\Sigma^+ \bar{\Sigma}^-$ and $\Sigma^0 \bar{\Sigma}^0$ decays from BESIII had a statistical significance of less than 5~$\sigma$ and were of limited precision. To make an objective comparison, the BF for the decays $\chi_{cJ} \rightarrow \Sigma \bar{\Sigma}$ from the previous BESIII publications are corrected with the newest BF of $\psi(3686) \rightarrow \gamma \chi_{cJ}$ from Ref.~\cite{PDG16}. To be independent of the BF of $\psi(3686) \rightarrow \gamma \chi_{cJ}$, the product BF ($\mathcal{B}_{\rm prod}$) of  $\psi(3686) \rightarrow \gamma \chi_{cJ}$ and $\chi_{cJ} \rightarrow \Sigma \bar{\Sigma}$ are also listed in Table~\ref{tab:finalres}. The ratios of the BF between $\chi_{cJ} \rightarrow \Sigma^+ \bar{\Sigma}^-$ and $\Sigma^0 \bar{\Sigma}^0$ are shown in Table~\ref{tab:ratiores}. The results are consistent with the expectation of isospin symmetry. 

\begin{table}[thbp]
  \caption[BF results]{The ratio of BF between $\chi_{cJ} \rightarrow \Sigma^+ \bar{\Sigma}^-$ and $\Sigma^0 \bar{\Sigma}^0$. The first uncertainty is statistical and the second systematic. The systematic uncertainties of the same sources are cancelled.}
  \label{tab:ratiores}
  \begin{tabular}{lccccc} \hline
   \hline      Channels    &  Ratio\\ 
    \hline   
                $\mathcal{B}(\chi_{c0} \rightarrow \Sigma^+ \bar{\Sigma}^-$)/$\mathcal{B}(\chi_{c0} \rightarrow \Sigma^0 \bar{\Sigma}^0$)~  & ~$1.06\pm0.07\pm0.08$  \ST \\
                 $\mathcal{B}(\chi_{c1} \rightarrow \Sigma^+ \bar{\Sigma}^-$)/$\mathcal{B}(\chi_{c1} \rightarrow \Sigma^0 \bar{\Sigma}^0$)~    &  ~$0.86\pm0.17\pm0.07$ \ST \\
                 $\mathcal{B}(\chi_{c2} \rightarrow \Sigma^+ \bar{\Sigma}^-$)/$\mathcal{B}(\chi_{c2} \rightarrow \Sigma^0 \bar{\Sigma}^0$)~   &  ~$0.90\pm0.21\pm0.10$     \ST \\ \hline
    \hline
  \end{tabular}
\end{table}

\begin{table*}[thbp]
  \caption[systematic uncertainty]{Summary of relative systematic uncertainties for the measurement of $\chi_{cJ} \rightarrow \Sigma^+ \bar{\Sigma}^-$ and $\Sigma^0 \bar{\Sigma}^0$ (\%) . }
  \label{tab:system}
  \begin{tabular}{lcccccc} \hline 
    \hline  \\    Sources   & $\chi_{c0} \rightarrow \Sigma^+ \bar{\Sigma}^-$   & $\chi_{c1} \rightarrow \Sigma^+ \bar{\Sigma}^-$  &  $\chi_{c2} \rightarrow \Sigma^+ \bar{\Sigma}^-$ & $\chi_{c0} \rightarrow \Sigma^0 \bar{\Sigma}^0$   & $\chi_{c1} \rightarrow \Sigma^0 \bar{\Sigma}^0$  & ~$\chi_{c2} \rightarrow \Sigma^0 \bar{\Sigma}^0$\\ 
    \hline      Number of $\psi(3686)$   &   0.6      &    0.6  &    0.6 &   0.6      &    0.6  &    0.6 \\
                   Photon selection  & 0.6 & 0.6 & 0.6 & 1.8 & 1.8 & 1.8 \\
                    Tracking and PID  & 2.7    & 2.7 & 2.7 & ---   & --- & --- \\ 
                     $\Lambda$ and $\bar{\Lambda}$ reconstruction & ---   & --- & --- & 4.5  & 4.5 & 4.5 \\
                    $\pi^0$ reconstruction & 2.4 & 2.4 & 2.4 & --- & --- & ---\\
                    $\Sigma$ mass window & 0.3 & 0.3 & 0.3 & 0.6  & 0.6 & 0.6\\
                    4C kinematic fit  & 0.8 & 0.8 & 0.8  & 3.1 & 3.1 & 3.1 \\
                    $\mathcal{B}$($\psi(3686) \rightarrow \gamma \chi_{cJ}$)  & 2.7 & 3.2 & 3.4 & 2.7 & 3.2 & 3.4 \\
                    $\mathcal{B}$($\Sigma^+ \rightarrow p \pi^0$) + c.c.  & 1.2 & 1.2 & 1.2 & ---   & --- & ---  \\
                    $\mathcal{B}$($\Lambda \rightarrow p \pi^-$)  + c.c. & ---   & --- & ---  & 1.6 & 1.6 & 1.6 \\
                    Fit range & 0.8 & 0.9 & 1.9 & 1.1 & 1.2 & 2.3 \\
                    $\chi_{c1, 2}$ width & --- & 0.1 & 0.3 & --- & 0.1 & 0.3 \\
                    Background shape  & 1.4 & 1.6 & 4.1 & 1.7 & 3.1 & 1.2 \\
                    Signal shape  & 1.5 & 2.3 & 3.7 & 1.7 & 0.9 & 1.5 \\
                    Peaking background & 0.9 & 3.1 & 5.0 & ---   & --- & --- \\ 
                    Generator  & 0.1 & 1.1 & 1.5 & 2.0   & 2.3 & 2.5 \\\hline
                    Total  & 5.4 & 6.8 & 9.4 & 7.4 & 8.0 & 8.0\\ \hline 
    \hline
  \end{tabular}
\end{table*}

\section{Systematic uncertainties}
\label{sec:systematic}
The systematic uncertainties are summarized in Table~\ref{tab:system}.  The number of $\psi(3686)$ events is determined by counting inclusive hadronic events from $\psi(3686)$ decays with an uncertainty of 0.6\% (see Ref.~\cite{Zhiyong17} for a description of the method). A control sample of $J/\psi \rightarrow \pi^+ \pi^- \pi^0$ decays is used to study the efficiency of the photon selection. The systematic uncertainty of the photon selection is estimated to be 0.5\% for the barrel and 1.5\% for the end-caps.  As a result, the systematic uncertainty from the photon selection efficiency in the present analysis is assigned to be 0.6\% per photon by means of a weighted average. In the decay $\chi_{cJ} \rightarrow \Sigma^+ \bar{\Sigma}^-$, only the radiative photon is considered for the uncertainty of photon detection.  The tracking and particle identification (PID) efficiencies of proton (anti-proton) from $\Sigma^+$ ($\bar{\Sigma}^-$) decay are studied using a control sample of $J/\psi \rightarrow \Sigma^+ \bar{\Lambda} \pi^- + c.c.$  The number of $\Sigma^+$ events with and without tracking and PID of the proton can be extracted from the distribution of the recoil mass of $\bar{\Lambda} \pi^-$, and the ratio of the corresponding numbers is assigned to be the detection efficiency. The difference of the tracking and PID efficiencies between data and MC samples is determined to be 1.3\% for protons and 1.4\% for anti-protons and is assigned as the systematic uncertainty.  The $\Lambda$ and $\bar{\Lambda}$ reconstruction efficiencies are studied using a control sample of $\psi(3686) \rightarrow \Lambda \bar{\Lambda}$ decays. The number of $\Lambda$ events before and after reconstruction can be extracted from the recoil mass of the $\bar{p} \pi^+$ and vice versa. The differences of the reconstruction efficiency between MC simulation and data, 2.0\% for $\Lambda$ and 2.5\% for $\bar{\Lambda}$, are assigned as the systematic uncertainty.  The systematic uncertainties from tracking and PID of charged tracks in the $\Lambda ~(\bar{\Lambda})$ decay are included in this number.  The systematic uncertainty due to the $\Lambda$ mass window cut is determined to be 0.2\%, and the requirement of the $\Lambda$ decay length to be greater than zero introduces a systematic uncertainty of 0.4\%. These two contributions to the systematic uncertainty are combined into the $\Lambda$ and $\bar{\Lambda}$ reconstruction uncertainty. The $\pi^0$ reconstruction efficiency is studied using control samples of $\psi(3686) \rightarrow J/\psi \pi^0 \pi^0$ and $\psi(3770) \rightarrow \omega \pi^0$ events, individually. The relative difference of the $\pi^0$ reconstruction efficiency (including the photon detection efficiency) between data and MC is found to be 1.2\% in both samples, which we assign as a systematic uncertainty. The   
$\pi^0$ mass window does not contribute significantly to the uncertainty.  The systematic uncertainty due to the $\Sigma^{+}$ and $\bar{\Sigma}^-$ ($\Sigma^0$ and $\bar{\Sigma}^0$ ) mass window cut is determined to be 0.3\% (0.6\%) using a control sample of $\psi(3686) \rightarrow \Sigma^+ \bar{\Sigma}^-$ ($J/\psi \rightarrow \Sigma^0 \bar{\Sigma}^0$) decay. 

The systematic uncertainty of the 4C kinematic fit for $\chi_{cJ} \rightarrow \Sigma^+ \bar{\Sigma}^-$ is studied using a control sample of $\psi(3686) \rightarrow \pi^0 \pi^0 J/\psi; J/\psi \rightarrow p \bar{p} \gamma$ decay by correcting the charged track helix parameters~\cite{helix}. The difference of 0.8\% in efficiency between the simulation and the data is assigned as the systematic uncertainty.  For the neutral mode $\chi_{cJ} \rightarrow \Sigma^0 \bar{\Sigma}^0$, we use control samples of $J/\psi \rightarrow \Sigma^0 \bar{\Sigma}^0$ and $\psi(3686) \rightarrow \pi^0 \pi^0 J/\psi; J/\psi \rightarrow p \bar{p} \pi^+ \pi^-$ events to estimate the systematic uncertainty due to the 4C kinematic fit. The larger difference of 3.1\% between MC and data is assigned as the 4C fit systematic uncertainty.  The systematic uncertainty of the decay BF of intermediate states is obtained from the uncertainties quoted in the PDG.  The uncertainty from the determination of $\chi_{cJ}$ events due to the fit range is obtained from the maximum difference in the fit result by changing the fit range from 3.3-3.6~GeV$/c^2$ to 3.25-3.6~GeV$/c^2$ or 3.25-3.61~GeV$/c^2$.  Since the number of $\chi_{c1, 2}$ events is small, the width of the $\chi_{c1, 2}$ signal shape is fixed to the PDG value. Changing the width within $\pm 1\sigma$ of the quoted uncertainty, the maximum difference is assigned as the systematic uncertainty. The systematic uncertainty due to the detector resolution is found to be negligible using the control sample of  $J/\psi \rightarrow \Sigma^+ \bar{\Sigma}^-$ and $J/\psi \rightarrow \Sigma^0 \bar{\Sigma}^0$. The shape of the background in the fit is changed from a second order Chebyshev polynomial to a first or third order one, individually, and the maximum difference in the fit result is assigned as the systematic uncertainty.  By changing the damping factor from $e^{-\frac{E_{\gamma}^2}{8\beta^2}}$ used by CLEO~\cite{CLEO}  to $\frac{E_0^2}{E_{\gamma}E_0+(E_{\gamma}-E_0)^2}$ used by KEDR~\cite{KEDR}, the differences in the fit results are  assigned as the systematic uncertainty due to the signal line shape.  The systematic uncertainty due to peaking background is obtained by changing the boundary of the sideband, the fit range, the shape of the background and signal in the sideband data similarly as described above as well as the scale factor $s$ of the MC simulation obtained from $\chi_{c0}$ to that obtained from $\chi_{c1}$, $\chi_{c2}$ decays and a uniform assumption ($s=1$).  The distribution of the polar angle of $\Sigma^+$ in the $\chi_{cJ}$ rest frame is used to study the angular distribution of $\chi_{cJ} \rightarrow \Sigma^+ \bar{\Sigma}^-$ decays.  The function $(1+\alpha \text{cos}^2\theta)$ is used to fit the data. Alternative signal MC samples are generated by changing the $\alpha$ value by $\pm 1\sigma$ of the fit value.  The resulting maximum difference in the efficiency is assigned as the systematic uncertainty. A similar systematic uncertainty is assigned to the neutral modes. By changing the weak decay parameters of the baryons within $\pm 1\sigma$ of the uncertainties quoted by the PDG, we find the resulting maximum difference in the detection efficiency to be 0.1\% and 2\% for the charged and neutral decay modes.  These two terms associated with modeling the decays are combined into the generator uncertainty. 
 The total systematic uncertainty is obtained by adding the individual uncertainties in quadrature. 
 
\section{SUMMARY}
\label{sec:summary}
In summary, using the world's largest $\psi(3686)$ sample at on-resonance production taken with the BESIII detector, we have measured the BF of $\chi_{cJ} \rightarrow \Sigma^+ \bar{\Sigma}^-$ and $\Sigma^0 \bar{\Sigma}^0$. The results presented replace the previous BESIII results~\cite{proton}. The decays $\chi_{c1, 2} \rightarrow \Sigma^+ \bar{\Sigma}^-$ and $\Sigma^0 \bar{\Sigma}^0$ are observed with more than $5\sigma$ significance for the first time.  The results are consistent with and improve on the precision compared to the world average values. The current results on $\chi_{c1, 2} \rightarrow \Sigma^+ \bar{\Sigma}^-$ and $\Sigma^0 \bar{\Sigma}^0$ are in good agreement with theoretical predictions based on the color octet contribution model~\cite{Wong00}. The results for $\chi_{c0} \rightarrow \Sigma^+ \bar{\Sigma}^-$ and $\Sigma^0 \bar{\Sigma}^0$ are still inconsistent with the prediction~\cite{Liu2011}  based on the charm meson loop mechanism. 
The ratio between charged and neutral decay modes is consistent with the expectation from isospin symmetry.  

\section{ACKNOWLEDGMENT}
The BESIII collaboration thanks the staff of BEPCII and the IHEP computing center for their strong support. This work is supported in part by National Key Basic Research Program of China under Contract No. 2015CB856700; National Natural Science Foundation of China (NSFC) under Contracts Nos. 11375204, 11505034, 11235011, 11335008, 11425524, 11625523, 11635010; the Chinese Academy of Sciences (CAS) Large-Scale Scientific Facility Program; the CAS Center for Excellence in Particle Physics (CCEPP); the Collaborative Innovation Center for Particles and Interactions (CICPI); Joint Large-Scale Scientific Facility Funds of the NSFC and CAS under Contracts Nos. U1332201, U1532257, U1532258; CAS Key Research Program of Frontier Sciences under Contracts Nos. QYZDJ-SSW-SLH003, QYZDJ-SSW-SLH040; 100 Talents Program of CAS; National 1000 Talents Program of China; INPAC and Shanghai Key Laboratory for Particle Physics and Cosmology; German Research Foundation DFG under Contracts Nos. Collaborative Research Center CRC 1044, FOR 2359; Istituto Nazionale di Fisica Nucleare, Italy; Koninklijke Nederlandse Akademie van Wetenschappen (KNAW) under Contract No. 530-4CDP03; Ministry of Development of Turkey under Contract No. DPT2006K-120470; National Natural Science Foundation of China (NSFC) under Contracts Nos. 11505034, 11575077; National Science and Technology fund; The Swedish Research Council; U. S. Department of Energy under Contracts Nos. DE-FG02-05ER41374, DE-SC-0010118, DE-SC-0010504, DE-SC-0012069; University of Groningen (RuG) and the Helmholtzzentrum fuer Schwerionenforschung GmbH (GSI), Darmstadt; WCU Program of National Research Foundation of Korea under Contract No. R32-2008-000-10155-0. 


\end{document}